\documentclass[a4paper,11pt]{article}

\usepackage{jheppub} 
\usepackage{subfigure}
\usepackage{graphicx}
\usepackage{amsfonts}

\preprint{KUNS-2996}

\title{\boldmath 
Neural network representation of quantum systems
}

\author[]{Koji Hashimoto,}
\author[]{Yuji Hirono,}
\author[]{Jun Maeda,}
\author[]{Jojiro Totsuka-Yoshinaka}

\affiliation[]{Department of Physics, Kyoto University, Kyoto 606-8502, Japan}

\emailAdd{koji@scphys.kyoto-u.ac.jp}
\emailAdd{yuji.hirono@gmail.com}
\emailAdd{maeda@gauge.scphys.kyoto-u.ac.jp}
\emailAdd{george.yoshinaka@gauge.scphys.kyoto-u.ac.jp}

\abstract{ 
It has been proposed that random wide neural networks near Gaussian process are quantum field theories around Gaussian fixed points. 
In this paper, we provide a novel map with which 
a wide class of quantum mechanical systems can be cast into the form of
a neural network with a statistical summation over network parameters. 
Our simple idea is to use the universal approximation theorem of neural networks
to generate arbitrary paths in the Feynman's path integral.
The map can be applied to interacting quantum systems / field theories,
even away from the Gaussian limit. 
Our findings bring machine learning closer to the quantum world.
}

\begin{document}

\maketitle
\flushbottom


\section{Introduction}
\label{sec:Intro}

Needless to be exemplified by Boltzmann machine, Amari-Hopfield model and
diffusion models, fundamental physics has provided a great influence on machine learning. Then a natural question arises --- to what extent do the fundamental physics and machine learning overlap with each other? For example, the notion of {\it quantum} is the central concept in microscopic physics. {\it To what extent can quantum mechanics be formulated in terms of neural networks?}

Partial answers to this interesting question come from two developments at the intersection of machine learning and physics: (1) Gaussian processes and (2) stochastic neurodynamics, which we shall describe in order.
Both of these have their roots in the research of 
random neural networks initiated by Amari~\cite{Amari,Amari2} and Rozonoer~\cite{Rozonoer}. The random neural network is a fundamental tool to reveal the macroscopic properties of typical neural networks, as well as a key to control complicated learning dynamics of neural networks. 

Recently, motivated by the findings that infinitely wide neural networks exhibit Gaussian processes \cite{Neal, GP, Greg} due to the central limit theorem for the random network parameters, 
controlling the learning dynamics can be systematically understood by the kernel expansion of the perturbation around the Gaussian process limit \cite{NTK}. A Feynman diagrammatic method was also developed \cite{Dyer:2019uzd}.
Along this development, the correspondence between 
wide neural networks and quantum field theories (QFTs) was proposed \cite{Halverson:2020trp,Halverson:2021aot,Demirtas:2023fir}, and is called  Neural Network Field Theory (NNFT). The NNFT chooses a specific activation function and probability density of network parameters to reproduce the free QFT correlators in the wide limit of the network. The NNFT serves a new representation of QFTs near the free QFTs.

Historically, randomness in neural networks has been assumed for the purpose of analysing the macroscopic dynamics of neurons, which leads to the statistical field theory techniques applied to stochastic time evolution of neural data \cite{Somp,Schucker,Helias}. Since it uses a path-integral formulation of the Langevin equation, a natural connection to QFTs was recently discussed \cite{Grosvenor:2021eol}. This direction reminds us of the stochastic quantization \cite{Nelson:1966sp,Parisi:1980ys,Damgaard:1987rr}
and diffusion models in machine learning \cite{diff,song2021scorebased}, and the relation between them was studied \cite{Wang:2023exq,Wang:2023sry}.

In this paper, we present a new neural network representation of quantum mechanics. 
Any quantum mechanical system can be formulated by 
Feynman's path-integral \cite{Feynman:1948ur,FeynmanHibbs}
in which 
arbitrary paths (configurations) of the quantum mechanical degree of freedom (field in the case of QFTs) are generated and summed.
In our approach, for the generation, we use the universal approximation theorem \cite{UAT1,UAT2,UAT3,UAT4,UAT5} for the neural networks which proves that any continuous function can be approximated as a neural network output. 
Identifying the network outputs as the paths for the path integral, we can represent quite wide class of quantum systems as a statistical average over the neural network outputs.

\begin{figure}[t]
\centerline{\includegraphics[width=0.8\columnwidth]{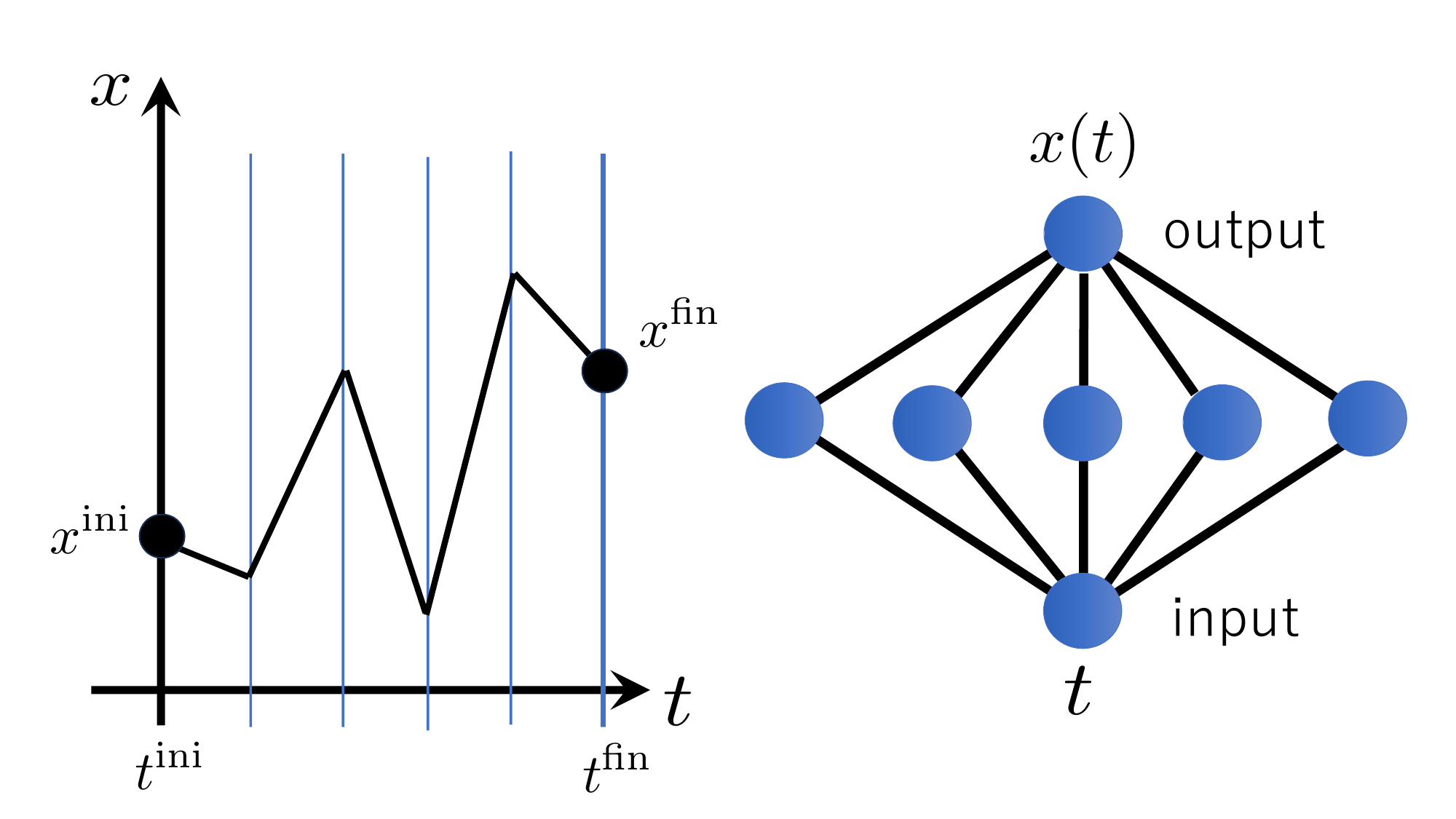}}
\caption{Left: a path of $x(t)$ in the quantum-mechanical path-integral. Right: a multilayer perceptron, our neural network architecture.}
\label{fig1}
\end{figure}

In the left panel of Fig.~\ref{fig1}, a generic path for the path integral in a single-variable quantum-mechanical system is shown. It is zigzag shaped, and arbitrary continuous paths are necessary for the path integral. Now, using the neural network shown in the right panel of Fig.~\ref{fig1}, this zigzag arbitrary path can be written as the output function $x(t)$, where the input is the time variable $t$, with the ReLU activation function. The arbitrariness of the path is assured by the universal approximation theorem. 
The integration measure of the path-integral, originally $\int {\cal D}x$, is replaced by a measure of the integration over the neural network parameters (known as network weights and biases) --- quantum mechanics is rewritten in terms of a certain statistical average of the neural networks whose parameters vary. 
The statistical weights for the network parameters are chosen such that they reproduce quantum mechanics. In this sense, our neural network is not completely random but subject to a certain statistical distribution.

Note that in our prescription we do not use any Gaussian process for the central limit theorem, nor the stochastic time evolution. Our method is direct reinterpretation of the path integral in any quantum mechanics and QFT. 
We note also that the NNFT and the stochastic quantization naturally appears in our scheme, since both of them can be reinterpreted in terms of path integrals.

The organization of this paper is as follows. In Sec.~\ref{sec:NNFT}, we give a brief review of the NNFT. It uses the same network architecture as ours while the statistical weights of the network parameters and the activation function are different. 
In Sec.~\ref{sec:QMRNN} we explain our method of representing an arbitrary path of the quantum path integral as a neural network output, by using the universal approximation theorem.
We explicitly provide a neural network representation of a free single particle, a harmonic oscillator and a $\phi^4$ theory as examples.
In Sec.~\ref{sec:ACT}, we use different activation functions for the quantum mechanics representation, and discuss the physical meaning of the neural network parameters. In particular, we clarify the relation of the present formulation to the NNFT.
In Sec.~\ref{sec:Euclid}, we provide a neural network representation of Euclidean quantum mechanics for wave functions. We discuss a relation to stochastic quantization.
The final section is for a summary and discussions.
In App.~\ref{app:A}, we discuss the gauge symmetry of neural networks to find our scheme quite generic, and 
in App.~\ref{app:B}, we study the detailed relation between our neural network representation and stochastic quantization.


\section{A brief review of the Neural Network Field Theory}
\label{sec:NNFT}

Our method is highly motivated by the Neural Network Field Theory (NNFT) \cite{Halverson:2020trp,Halverson:2021aot,Demirtas:2023fir},
so here in this section we provide a short review of it.
In particular, the same network architecture of the NNFT is shared with ours, but the statistical distribution of the network parameters and the activation function are different. 
We focus on the quantum-mechanics version of the NNFT, for the purpose of comparison with our approach, which is described in Sec.~\ref{sec:Euclid}.

The essential idea of the NNFT is that, when the width $N$ of a neural network is infinite ($N\to\infty$) and
the network parameters are drawn from a unique probability distribution, {\it i.e.} the parameters are i.i.d.~samples, 
then the central limit theorem applies to the output of the neural network, resulting in a Gaussian process.
In terms of QFTs, the Gaussian process means that all possible correlators are given
just by Wick contraction for pairs, reducing to a sum of products of the two-point functions. 
This feature is well-known to exist for free QFTs. Then, the remaining task to concretely 
relate the neural network output with a quantum field is to reproduce the QFT two-point function by tuning the activation function and the statistical distributions of the neural network parameters, which is 
realized as follows.

We shall see here the NNFT construction of a (Euclidean) quantum harmonic oscillator.
One starts with the same neural network architecture shown in the right panel of Fig.~\ref{fig1}, 
and considers the cosine-type activation with a particular normalization,
\begin{eqnarray}
    x(\tau) = \sum_{n=1}^N a_n \frac{1}{\sqrt{b_n^2 + k/m}} \cos\left(b_n \tau + c_n\right) \, .
    \label{NNFTQM}
\end{eqnarray}
Here $\{a_n, b_n, c_n\}$ are the neural network parameters, and $k$ and $m$ are for the harmonic oscillator,
whose Euclidean Lagrangian is $m \dot{x}^2 + k x^2$.
$\{b_n, c_n\}$ are the network parameters for the network connecting the single unit in the first layer (whose input is the Euclidean time $\tau$) and the $N$ units in the second layer. $\{a_n\}$ are the network parameters connecting the second layer and the third layer. The final output at the third layer is named $x(\tau)$.

To apply the central limit theorem, one assumes that 
the probability distribution of these parameters do not depend on $n$, 
in other words, $\{a_n, b_n, c_n\}$ are i.i.d. samples.
And one introduces the following probability distribution functions,
\begin{eqnarray}
    & a_n \sim P(a) = {\cal N}(0,1/N)\, , \\
    & b_n \sim P(b) = {\cal U}(-\Lambda, \Lambda) \, ,\\
    & c_n \sim P(c) = {\cal U}(-\pi, \pi) \, .
\end{eqnarray}
Here ${\cal N}(0,1/N)$ is the normal distribution with the vanishing mean and the variance $1/\sqrt{N}$, 
and ${\cal U}(-\alpha, \alpha)$
is a uniform distribution. $\Lambda$ is a cut-off parameter which will be taken to $\infty$ when required.
Using these, one can show that the two-point function is evaluated as
\begin{eqnarray}
    \langle x(\tau_1) x(\tau_2)\rangle
& = &\int \prod_n (P(a_n)da_n P(b_n) db_n P(c_n)dc_n) \, 
x(\tau_1)x(\tau_2)
\nonumber \\
& =  & \displaystyle
\int_{-\Lambda}^\Lambda db \, \frac{\cos(b(\tau_1-\tau_2))}{b^2 + k/m} \, .
\end{eqnarray}
This is a Fourier transform of the propagator $1/(b^2+k/m)$, and the integrated parameter $b$ is 
conjugate to the time variable. 
Thus in the NNFT construction, the interpretation of the network parameters
is that 
the parameter $b$ is a frequency (or the temporal component of the momentum), 
and the parameter $a$ is the Fourier coefficient. 

This illustration of a quantum mechanics is generalized to free quantum field theories in $d$ spacetime dimensions, 
by preparing $d$ units at the first layer which take values of the spacetime coordinates.
In this NNFT, the realization of free field theories is tied to the central limit theorem and the Gaussian process. To add interactions, one needs to go away from the Gaussian process. 
Two strategies were proposed: (1) going to the finite $N$ regime, then the $1/N$ corrections amounts to an interaction of fields \cite{Halverson:2020trp,Halverson:2021aot,Demirtas:2023fir}, (2) breaking of the i.i.d.~property, by providing a multiplicative factor in the probability distribution function to reproduce given interaction terms in the QFT Lagrangian \cite{Demirtas:2023fir}.


\section{Quantum mechanics as a neural network}
\label{sec:QMRNN}

In this section, we provide our main idea on our neural network representation of quantum systems (Sec.~\ref{sec:3.1}), followed by explicit examples of a free particle (Sec.~\ref{FP}), a harmonic oscillator (Sec.~\ref{HO}) and quantum field theories (Sec.~\ref{sec:QFT}).

\subsection{Path integral and statistical summation of network parameters}
\label{sec:3.1}

We first describe how a quantum path-integral can be rewritten as a neural network.
For simplicity, we consider a particle moving in one dimension, whose location is specified as 
$x(t)$ at time $t$.
The standard definition of a transition amplitude for a particle with use of a path integral is
\begin{eqnarray}
\langle x^{\rm fin} | x^{\rm ini} \rangle
= \int 
\left[\prod_{n=1}^{N-1} d x_n\right] \, 
\exp\left[\frac{i}{\hbar}S[x]\right]
\label{tra}
\end{eqnarray}
where $x_n \equiv x(t=t_n)$, with the period of time $t^{\rm ini}\leq t \leq t^{\rm fin}$ divided into $N$ parts
as 
\begin{eqnarray}
t_n \equiv t^{\rm ini} + n \Delta t \quad (n=0,1,\cdots,N) \, , \quad 
\Delta t \equiv \frac{t^{\rm fin}-t^{\rm ini}}{N} \, .
\end{eqnarray}
This means $t^{\rm ini} = t_0$ and $t^{\rm fin}=t_N$, and $x^{\rm ini} = x_0$ and $x^{\rm fin}=x_N$. The expression (\ref{tra})
is for the amplitude with the initial condition of the particle sitting at $x=x^{\rm ini}$
at time $t=t^{\rm ini}$ and the final condition $x=x^{\rm fin}$ at $t=t^{\rm fin}$.
The action in (\ref{tra}) is given with a Lagrangian ${\cal L}$, which is discretized as
\begin{eqnarray}
S = \int dt \, {\cal L}[x,\dot{x}] = \Delta t \sum_{n=1}^{N} {\cal L}
\left[\frac{x_n+x_{n-1}}{2}, \frac{x_n-x_{n-1}}{\Delta t}\right] \, .
\label{midpoint}
\end{eqnarray}
The path integral says that one needs to make an integration over all possible configuration of $x(t)$ satisfying the boundary condition $x(t^{\rm ini})=x^{\rm ini}$ and $x(t^{\rm fin})=x^{\rm fin}$. Once the time coordinate is discretized, the path-integral is regularized by $N$ and one makes the integration over all possible zigzag configurations of ${x_n}$.
See Fig. \ref{fig1}.

Now, let us provide a neural network representation of the zigzag configuration $x(t)$.  Using the activation function ReLU$(x) \equiv {\rm max}(0,x)$, we consider a feedforward neural network whose input and output are one-dimension while at the unique hidden layer we have $N$ units. The input value is time $t$, and the output value is regarded as $x(t)$. The architecture shows  
\begin{eqnarray}
x(t) = \sum_{n=1}^{N} w_n {\rm ReLU}(\tilde{w}_n t + \tilde{b}_n) + b \, ,
\label{xtrelu}
\end{eqnarray}
where $w_n, b, \tilde{w}_n, \tilde{b}_n$ are the network parameters. 
Once we put a constraint on the parameters for the first layer as
\begin{eqnarray}
\tilde{w}_n = 1, \quad \tilde{b}_n = -t_{n-1} \, ,
\label{tilfix}
\end{eqnarray}
then the neural network output is 
\begin{eqnarray}
x(t) = \sum_{n=1}^{N} w_n {\rm ReLU}( t -t_{n-1}) + b \, ,
\label{xtrelu2}
\end{eqnarray}
which is found to be exactly of the zigzag form for the path-integral shown in Fig.~\ref{fig1}, through the identification
\begin{eqnarray}
x_n = x^{\rm ini} + \Delta t\sum_{k=0}^n k\, w_{n-k+1}   \, ,
\label{xn}
\end{eqnarray}
with $b=x^{\rm ini}$. 
The path-integral variables are $N-1$ variables $(x_1, x_2, \cdots, x_{N-1})$, 
while the neural network has $N$ variables $(w_1, w_2, \cdots, w_N)$. In fact, 
by the use of (\ref{xn}), the last parameter
$w_N$ is going to be fixed by the path-integral boundary condition at $x=x^{\rm fin}$ as
\begin{eqnarray}
w_N = \frac{1}{\Delta t}(x^{\rm fin} -x^{\rm ini}) - \sum_{k=2}^N k w_{N-k+1} \,  .
\label{wn}
\end{eqnarray}
In this way, the path-integral zigzag form of the function $x(t)$ in Fig.~\ref{fig1} has the neural network representation.

The important issue in this interpretation of the path-integral by a statistical average of  neural networks is
the integration measure. Let us rewrite the standard path-integral measure by the parameters $\{w_n\}$. Using (\ref{xn}), we can easily find
\begin{eqnarray}
\prod_{n=1}^{N-1} dx_n = (\Delta t)^{N-1} \prod_{n=1}^{N-1} dw_n \, ,  
\end{eqnarray}
So the path-integral measure in quantum mechanics is just equal to be the product measure of the network parameters.

When we include $w_N$ in the measure, we incorporate the constraint equation (\ref{wn})
as 
\begin{eqnarray}
\prod_{n=1}^{N-1} dx_n = 
(\Delta t)^{N-1} 
\delta\left(
\sum_{k=1}^N k w_{N-k+1} - \frac{1}{\Delta t}(x^{\rm fin} -x^{\rm ini}) 
\right)
\prod_{n=1}^{N} dw_n \, . 
\end{eqnarray}
Then the whole path-integral is mapped to a statistical average of neural networks,
\begin{eqnarray}
\langle x^{\rm fin} | x^{\rm ini} \rangle
=
(\Delta t)^{N-1} \int
\delta\left(
\sum_{k=1}^N k w_{N-k+1} - \frac{1}{\Delta t}(x^{\rm fin} -x^{\rm ini}) 
\right)
\prod_{n=1}^{N} dw_n
\exp\left[\frac{i}{\hbar}S\right] \, , 
\end{eqnarray}
On the right hand side, 
the path-integral weight is given by $\exp[\frac{i}{\hbar}S]$ in which the 
action $S$ is written in terms of the network parameters as 
\begin{eqnarray}
S = 
\Delta t \sum_{n=1}^{N} {\cal L}
\left[x^{\rm ini} + \Delta t\sum_{k=1}^n \left(k-\frac12\right) w_{n-k+1} \,  , 
\sum_{k=1}^n w_k
\right] \,  \, .
\end{eqnarray}
This is obtained by the substitution of (\ref{xn}) into (\ref{midpoint}).
In this manner, we can map the path-integral expression of the 
quantum transition amplitude of a particle to a statistical averaging over 
neural networks with a certain architecture, with the statistical average on the 
neural network parameters given by the path-integral weight
$\exp[\frac{i}{\hbar}S]$. 

The path integral makes sense in the limit $N\to\infty$, and it 
corresponds to the large width limit of the neural network. 
So, in this mapping, only the infinite-width neural networks are allowed to have a physical
interpretation.

Note that, although the statistical weight for the neural network parameters is complex $\exp[\frac{i}{\hbar}S]$ in the description above, one can go to the Euclidean formulation of path integral in which we can replace the phase factor $\exp[\frac{i}{\hbar}S]$ by a real factor $\exp[-\frac{1}{\hbar}S_{\rm E}]$ where
$S_{\rm E}$ is the Euclidean action. We will study the situation with Euclidean time in Sec.~\ref{sec:Euclid}.

In our identification of the parameter representation (\ref{xtrelu}) and the path of $x(t)$ uses the fixed choice of some of the parameters as shown in (\ref{tilfix}). We can relax this condition so that the fixed choice is not necessary, by using a symmetry of generic neural networks and an arbitrariness of time divisions in path integrals. The details are presented in App.~\ref{app:A}.


\subsection{Example 1 : Free particle}
\label{FP}

We explicitly derive the neural network representation of a nonrelativistic free particle,
\begin{eqnarray}
{\cal L} = \frac{m}{2} \dot{x}^2 \, .
\end{eqnarray}
Discretizing the time direction and substituting (\ref{xn}), we find
\begin{eqnarray}
S = \frac{m \Delta t}{2}
\sum_{n=1}^{N}\left(\sum_{k=1}^n w_k\right)^2 \, .
\end{eqnarray}
Therefore the path-integral weight is Gaussian.

To make the path-integral easier to read, we make the following linear redefinition of the network parameters,
\begin{eqnarray}
W_n \equiv \sum_{k=1}^n w_k \, ,
\label{Ww}
\end{eqnarray}
with which the measure keeps the product form,
$\prod d w_n = \prod d W_n$.
Then the transition amplitude is written as
\begin{eqnarray}
\langle x^{\rm fin} | x^{\rm ini} \rangle
=
(\Delta t)^{N-1} \int 
\exp\left[\frac{i}{\hbar}S\right]
\prod_{n=1}^{N-1} dW_n
\end{eqnarray}
where 
\begin{eqnarray}
S \equiv 
\frac{m\Delta t}{2}
\left(
\sum_{n=1}^{N-1}\left(W_n\right)^2
+
\left(
\frac{x^{\rm fin}-x^{\rm ini}}{\Delta t}
-\sum_{k=1}^{N-1} W_k
\right)^2
\right) \, .
\end{eqnarray}
The expression is an independent set of Gaussian weights, except for the 
last term in the action which is for the physical boundary condition at $t=t^{\rm ini}$
and $t=t^{\rm fin}$.
%
%
%
One can interpret the last term in the action as a loss term of the neural network,
to make the average
of the network parameters $\{W_n\}$ to be equal to the average velocity of the particle --- the classical trajectory
of the particle is a uniform motion.
The meaning of the neural network parameters in our quantum mechanical representation will be further explored in the next section.


\subsection{Example 2 : Harmonic oscillator}
\label{HO}

We can add some potential terms for the particle motion,
\begin{eqnarray}
{\cal L} = \frac{m}{2} \dot{x}^2 - V(x)
\end{eqnarray}
and the harmonic oscillator is with  $V(x)\equiv \frac12 k x^2$.
We can substitute (\ref{xn}) also to the part of the potential and find 
\begin{align} 
S =
&
\frac{m \Delta t}{2}
\left(
\sum_{n=1}^{N-1}\left(W_n\right)^2
+
\left(
\frac{x^{\rm fin}-x^{\rm ini}}{\Delta t}
-\sum_{k=1}^{N-1} W_k
\right)^2
\right)
\nonumber \\
&
-\frac{k\Delta t }{2}
\sum_{n=1}^{N-1}\left(x^{\rm ini}-\frac12 \Delta t \, W_n +\Delta t \sum_{k=1}^n W_k\right)^2
 \, .
\end{align}
Again, this is an independent set of Gaussian weights, 
after a proper linear redefinition of the variables $\{ W_n\}$.

We immediately notice that if we include an unharmonic term, such as $x^4$,
then the network parameters can no longer be diagonalized. This means that the parameters are not independent from each other, and are not Gaussian. This is consistent with the central limit theorem which assumes the independence of each probability distribution to get the Gaussian 
distribution in the large $N$ limit. 

But the point here is that, whatever the potential is, we can explicitly 
construct the coupled parameter distribution, for the statistical neural networks to represent the given quantum mechanical system.
For example, for the potential with the $x^4$ term, as
$V(x) = \frac{1}{2}k x^2 + g x^4$, we find the statistical weight
$\exp\left[\frac{i}{\hbar}S\right]$
given by 
\begin{align}
S =
&\frac{m\Delta t}{2}
\left(
\sum_{n=1}^{N-1}\left(W_n\right)^2
+
\left(
\frac{x^{\rm fin}-x^{\rm ini}}{\Delta t}
-\sum_{k=1}^{N-1} W_k
\right)^2
\right)
\nonumber \\
&
-\frac{k\Delta t }{2}
\sum_{n=1}^{N-1}\left(x^{\rm ini} -\frac12 \Delta t \, W_n+ \Delta t \sum_{k=1}^n W_k\right)^2
\nonumber \\
&
-g \Delta t 
\sum_{n=1}^{N-1}\left(x^{\rm ini} -\frac12 \Delta t \, W_n+ \Delta t \sum_{k=1}^n W_k\right)^4
 \, .
\end{align}
For small $g$, we can Taylor-expand $\exp\left[\frac{i}{\hbar}S\right]$
perturbatively in $g$,
and evaluate the deviation away from the independent case.

\subsection{Example 3 : Quantum field theory}
\label{sec:QFT}

\begin{figure}[t]
\centerline{\includegraphics[width=0.8\columnwidth]{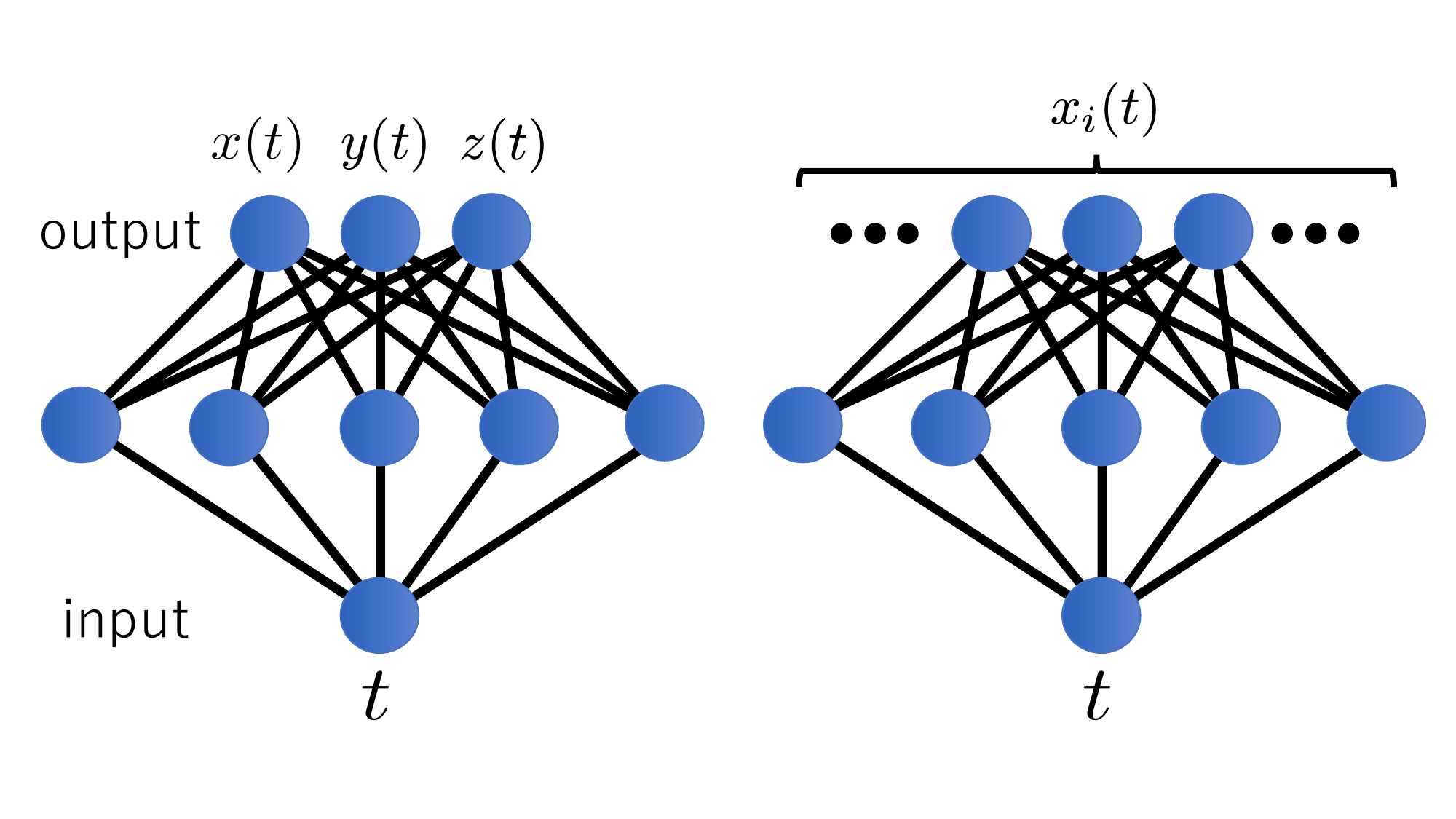}}
\caption{Left: a multilayer perceptron representing a quantum mechanics of a particle in three dimensional space. Right: A quantum field theory.}
\label{fig2}
\end{figure}

So far, we have given the neural network representation of the quantum mechanical systems in one dimension. 
Its generalization to higher dimensions is straightforward, see Fig.~\ref{fig2}.
We replace the output layer by that with multiple neurons whose number is the number of the degrees of freedom of the quantum mechanics. Precisely the same argument follows, with the identification given in
(\ref{xtrelu}) for each of the spatial coordinate.

In addition, 
if one takes the infinite limit of the number of neurons at the last layer, 
it can represent a scalar quantum field theory.
In fact, in the standard path-integral quantization of a quantum field theory, the path-integral variables are just $\phi({\mathbf x}, t_n)$ $(n=1,2, \cdots, N-1)$. Discretizing the space spanned by ${\mathbf x}$, we can rename the path-integral variables as $\phi_i(t_n)$ where $i$ $(=1,2, \cdots)$ labels the lattice site of the space. This is the quantum field theory on a lattice. Interpreting this $\phi_i(t_n)$ as $x_i(t_n)$ reduces everything
to the case of the multi-variate quantum mechanics. See Fig.~\ref{fig2}.

The $\phi^4$ scalar field theory Lagrangian in $d+1$-dimensions is
\begin{eqnarray}
{\cal L} = \int d^dx \left[\frac{1}{2} (\dot{\phi}(x,t))^2 - \frac12 (\nabla \phi(x,t))^2
-\frac{m^2}{2}\phi(x,t)^2-g\phi(x,t)^4\right]\, .
\end{eqnarray}
To reducing it to an infinite-dimensional quantum mechanics, we discretize the $d$-dimensional spatial direction. For simplicity we consider the case $d=1$ and introduce an equally-spaced lattice in which
the lattice constant is $\Delta x$ and the lattice sites are labeled by an integer $i\in {\mathbb Z}$. Then the Lagrangian is
\begin{eqnarray}
{\cal L} = \sum_{i=-\infty}^\infty
\left[\frac{1}{2} (\dot{\phi}_i(t))^2 - \frac{(\phi_{i+1}(t)-\phi_i(t))^2}{2(\Delta x)^2} 
-\frac{m^2}{2}\phi_i(t)^2 - g\phi_i(t)^4\right]\, .
\end{eqnarray}
We introduce a neural network of the architecture given in the right panel of Fig.~\ref{fig2}, 
\begin{eqnarray}
\phi_i(t) = \sum_{n=1}^{N} w^{(i)}_n {\rm ReLU}(t -t_{n-1}) + \phi^{\rm ini}_i \, .
\label{phitrelu}
\end{eqnarray}
This amounts to the zigzag configuration of the scalar field as
\begin{eqnarray}
\phi_{i,n} = \phi^{\rm ini}_i + \sum_{k=0}^n k w_{n-k+1}^{(i)} \Delta t  \, .
\label{phin2}
\end{eqnarray}
Then this map defines the path integral, under the identification of the statistical weight for the network parameters
with the action. Explicitly, the path-integral weight $\exp\left[\frac{i}{\hbar}S\right]$ is 
given by the following action
\begin{align}
    S = 
    &\Delta t \sum_{n=1}^{N}
    \sum_{i=-\infty}^\infty
\left[\frac{1}{2} \frac{(\phi_{i,n}-\phi_{i,n-1})^2}{(\Delta t)^2} - 
\frac{(\phi_{i+1,n}+\phi_{i+1,n-1}-\phi_{i,n}-\phi_{i,n-1})^2}{8(\Delta x)^2} 
\right.
\nonumber \\
&
\left.
\hspace{30mm}
-\frac{m^2}{8}(\phi_{i,n}+\phi_{i,n-1})^2 - \frac{g}{16}(\phi_{i,n}+\phi_{i,n-1})^4\right] \, ,
\end{align}
with the substitution of the expression (\ref{phin2}). Again, for $g=0$ everything is quadratic in the network parameters, 
meaning that this is a Gaussian distribution after a proper diagonalization. 
For the interacting quantum field theory with $g>0$, the statistical weight is not Gaussian.

A concluding important comment is that, we may not need a Gaussian limit for our neural network formulation of quantum systems.
Since we are using the path-integral representation of the quantum systems, it quantizes any action. For example, we can start with spin systems, as they allow coherent state path integral for their quantization, although they don't have Gaussian limit of the Lagrangian. The same is true for generic interacting quantum field theories whose low energy limit are not the trivial Gaussian limit. This shows the generality of our approach.


\section{Various activation functions and representation of quantum systems}
\label{sec:ACT}

\subsection{Step function activation}

\begin{figure}[t]
\centerline{\includegraphics[width=0.8\columnwidth]{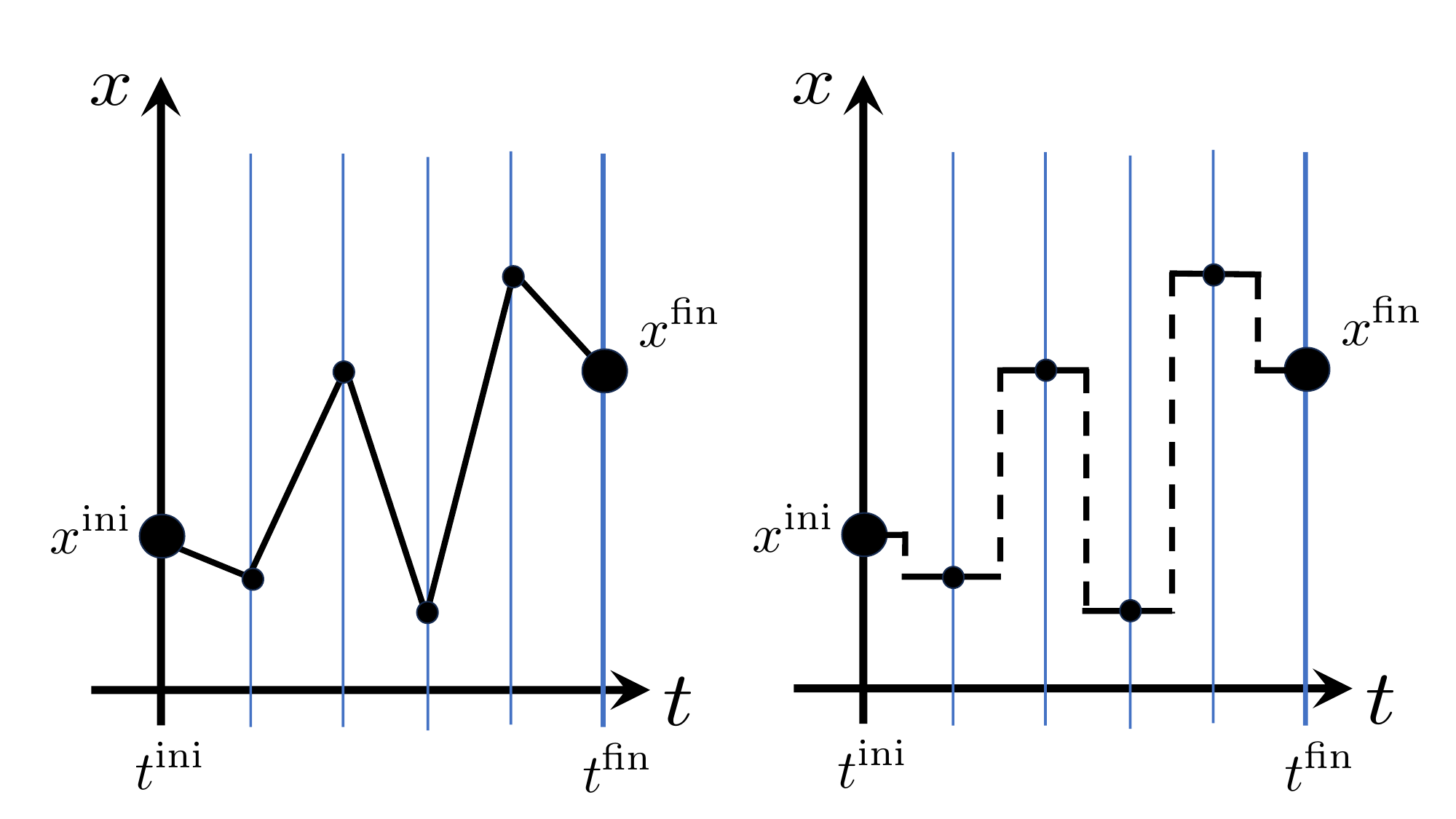}}
\caption{Left: a path of $x(t)$ using ReLU as its activation. Right: a path using a step function.}
\label{fig:ReLUstep}
\end{figure}

In generic neural networks, one can employ various activation functions, rather than ReLU used in Sec.~\ref{sec:QMRNN}. 
As an illuminating example 
which allows for 
more natural quantum-mechanical interpretation, let us use
a step function $\theta(t)$ as the activation function. 
This function can be obtained as a limit of the sigmoid function, which is commonly used in machine learning. The difference between the ReLU and the step function is illustrated
in Fig.~\ref{fig:ReLUstep}.

The output of the neural network architecture is
\begin{eqnarray}
    x(t) = x^{\rm ini} + \Delta t\sum_{n=1}^N W_n \, \theta\left(t-t_{n-1}-\frac{1}{2}\right)\, .
    \label{xstep}
\end{eqnarray}
Here the network parameters are $(W_1, W_2, \cdots, W_N)$ (scaled with the factor $\Delta t$). 
Notice that the network bias of the first affine map is shifted by $1/2$ so that the position of the path-integrated coordinate $x_n=x(t=t_n)$ is well-defined.
Using this, we can express the path-integral configuration $x_n$ as
\begin{eqnarray}
    x_n = x^{\rm ini} + \Delta t \sum_{k=1}^n W_k \, .
    \label{xnW}
\end{eqnarray}
As before, the last parameter $W_N$ needs to be fixed by the boundary condition, 
\begin{eqnarray}
    W_N=\frac{1}{\Delta t}(x^{\rm fin}-x^{\rm ini}) - \sum_{k=1}^{N-1}W_k \, .
\end{eqnarray}

We find that these are exactly the same expression as those of Sec.~\ref{sec:QMRNN}, using the definition
of $W_n$ given in (\ref{Ww}). In fact, the activation ReLU changes the slope of $x$ while the step function changes
the value of $x$ itself, so the relation between the parameters of those two different networks should be that of an integral over time, that is, the summation over the index $n$.
This means that the formulas for the statistical weights for the parameters, derived in Sec.~\ref{FP} and Sec.~\ref{HO},
still hold as they are, for this case of the neural network with the step function activation.

Through this argument, we find out the meaning of the parameters of the neural network. 
We can write (\ref{xnW}) as
\begin{eqnarray}
    W_n = \frac{x_n-x_{n-1}}{\Delta t} \, ,
    \label{xnW2}
\end{eqnarray}
and also (\ref{xn}) as
\begin{eqnarray}
    w_n = \frac{x_n-2x_{n-1}+x_{n-2}}{\Delta t} \, .
    \label{xnw2}
\end{eqnarray}
Therefore, $w_n$ is the acceleration of the particle motion at time $t=t_n$ and $W_n$ is the velocity at time $t=t_n$.
We were just rewriting the path integral in quantum mechanics, usually written by the integral over the coordinate $x_n$, by the integral over the velocity variable or the acceleration variable.


\subsection{Cosine activation and relation to the NNFT}
\label{subsec:Cos}

Here let us study the similarity and the difference between our map and  
the NNFT which we briefly reviewed in Sec.~\ref{sec:NNFT}.
Although the architecture and also the large $N$ limit are shared with our approach,
there are concrete differences and also
philosophical differences. In this subsection we clarify the difference,
through the viewpoint of the meaning of the network parameters --- 
in fact, the NNFT uses cosine-type activation to reproduce the 
ordinary kinetic term of a scalar field theory, while we use the ReLU or the step function
to reproduce the quantum system. More differences will be discussed in the final section.

\begin{table}
\begin{center}
\begin{tabular}{c|ccc}
& \multicolumn{3}{c}{Meaning of network parameters}\\
Activation & $a$ & $b$ & $c$\\\hline\hline
ReLU & Acceleration & Fixed & Time divisions (fixed)\\\hline
Step & Velocity & Fixed & Time divisions (fixed)\\\hline
Cosine & Fourier Coefficient & Frequency (fixed) & Fixed \\
\end{tabular}
\caption{
\label{tab1}
Meaning of the neural network parameters in our various architecture in the mapping to quantum systems. 
The neural network output is written as $x(t) = \sum_n a_n \sigma(b_n t + c_n)$, where $\sigma$ is the activation function.}
\end{center}
\end{table}

It is not apparent that the NNFT formulation itself is equivalent to the path-integral quantization, 
as the main argument of the NNFT is the use of the central limit theorem which states that
the large $N$ limit produces a Gaussian process.
In the following we clarify this point, by reminding the readers of the Fourier-basis path-integral.
First, in the ordinary path-integral of quantum mechanics of a harmonic oscillator,
we introduce a Fourier basis,
\begin{eqnarray}
    x(t) = \sum_n a_n \cos\left(\frac{n \pi}{T} t\right)\, . 
    \label{FTx}
\end{eqnarray}
Then the harmonic oscillator action is written as
\begin{eqnarray}
    S = \sum_n a_n^2 \left(
    m \left(\frac{n \pi}{T} \right)^2 - k 
    \right) \, .
\end{eqnarray}
The measure of the path-integral is 
\begin{eqnarray}
    \prod_n dx_n = J \prod_n d a_n
\end{eqnarray}
where $J$ is a dynamics-independent constant, since the Fourier transform is just a universal linear transformation.
With these, the two-point function is given as
\begin{eqnarray}
    \langle x(t_1) x(t_2)\rangle
& = \int J \prod_n da_n  \, 
x(t_1)x(t_2) \exp\left[\frac{i}{\hbar} \sum_n a_n^2 \left(
    m \left(\frac{n \pi}{T} \right)^2 - k 
    \right)\right]\, .
    \label{FTxx}
\end{eqnarray}
Upon Euclideanization, we can find a strucural similarity to the NNFT. The only difference is the origin of the propagator; (\ref{NNFTQM}) and (\ref{FTx}) are similar, except for the factor
$1/\sqrt{b^2+k/m}$ in (\ref{NNFTQM}), and in (\ref{FTxx}) the propagator comes from the integrals 
of the Gaussian $\{a_n\}$ which are not i.i.d.~samples.

\begin{figure}[t]
\centerline{\includegraphics[width=1\columnwidth]{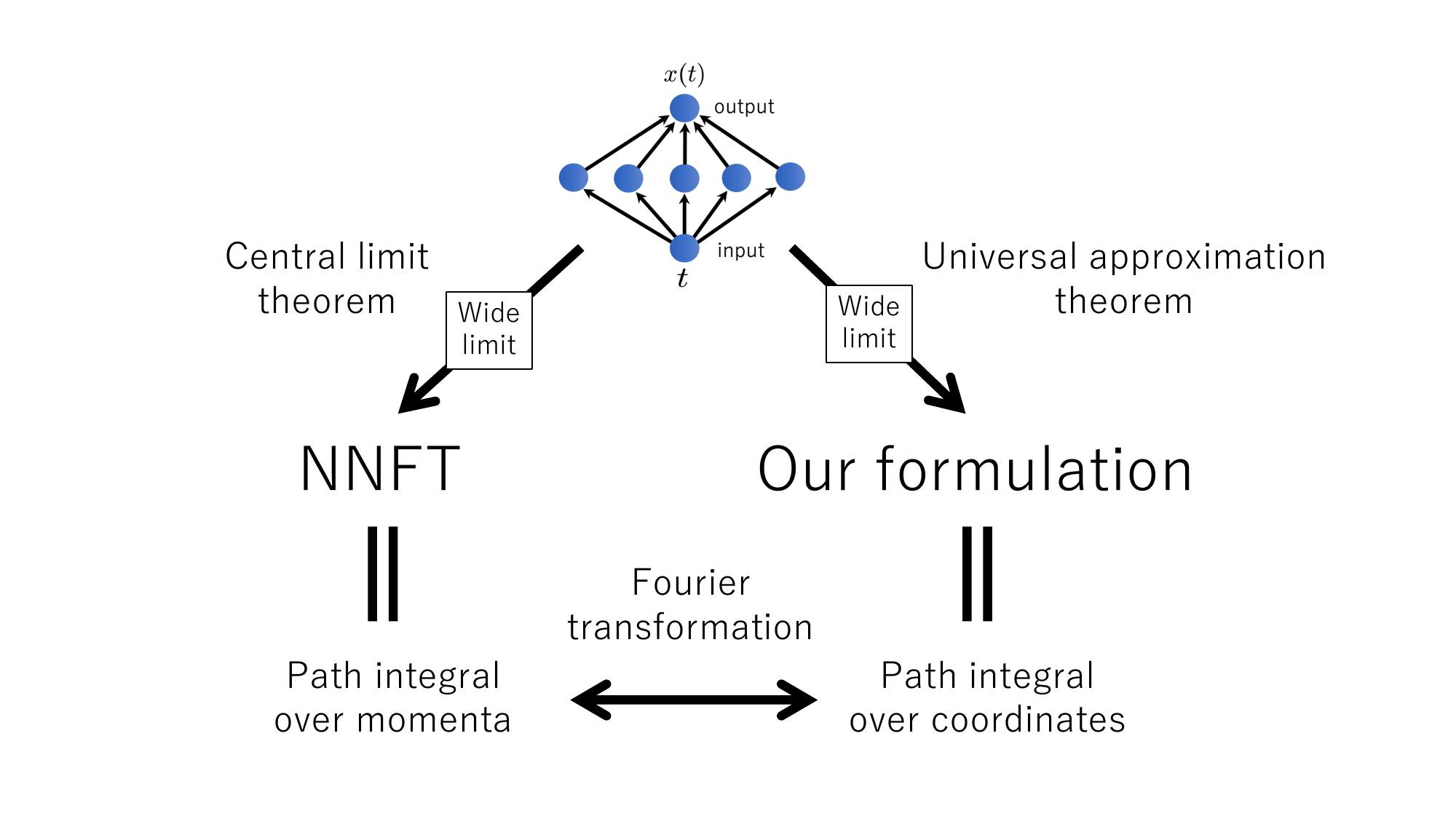}}
\vspace{-5mm}
\caption{A schematic picture of the relation between the NNFT and our formulation.}
\label{fig6}
\end{figure}

One conclusion here is that, if we regard (\ref{FTx}) as a neural network (with the cosine-type activation), 
then adjusting the neural network parameter distribution, we can obtain a quantum mechanics.
The NNFT is a well-arranged method in which the i.i.d.~property of the randomness is also maintained to reach the
central limit theorem to have the Gaussian process. On the other hand, 
as we have shown, 
in general the i.i.d.~property can be discarded, 
but still we can find a quantum mechanical neural network through the path-integral identification.
In Table \ref{tab1}, we summarize the meaning of the network parameters in our correspondence to quantum mechanics.
In Fig.~\ref{fig6}, we summarize the relation between the NNFT and our formulation.

\section{Quantum mechanics with Euclidean time as a neural network}
\label{sec:Euclid}

So far, we have explored the Lorentzian signature for the spacetime, with which the 
``probability distribution" 
$\exp\left[\frac{i}{\hbar}S\right]$
is a complex number. To make it real, we turn to the Euclidean formulation of quantum mechanics. 
We obtain the distribution of the neural network parameters, to represent the ground state wave function.
In addition, we provide the interpretation of the network parameters as a random walk through the stochastic quantization.

\subsection{Ground state wave function}

The Euclidean formulation of quantum mechanics provides a real-valued probability density of the network parameters, with which our mapping is more rigorously justified.
With the Euclidean time $\tau = it$, the Euclideanized Lagrangian of a quantum mechanics is
\begin{eqnarray}
{\cal L}_{\rm E}=\left[
\frac{m}{2} \left(\frac{dx}{d\tau}\right)^2 + V(x)
\right] \, .
\end{eqnarray}

\begin{figure}[t]
\centerline{\includegraphics[width=0.8\columnwidth]{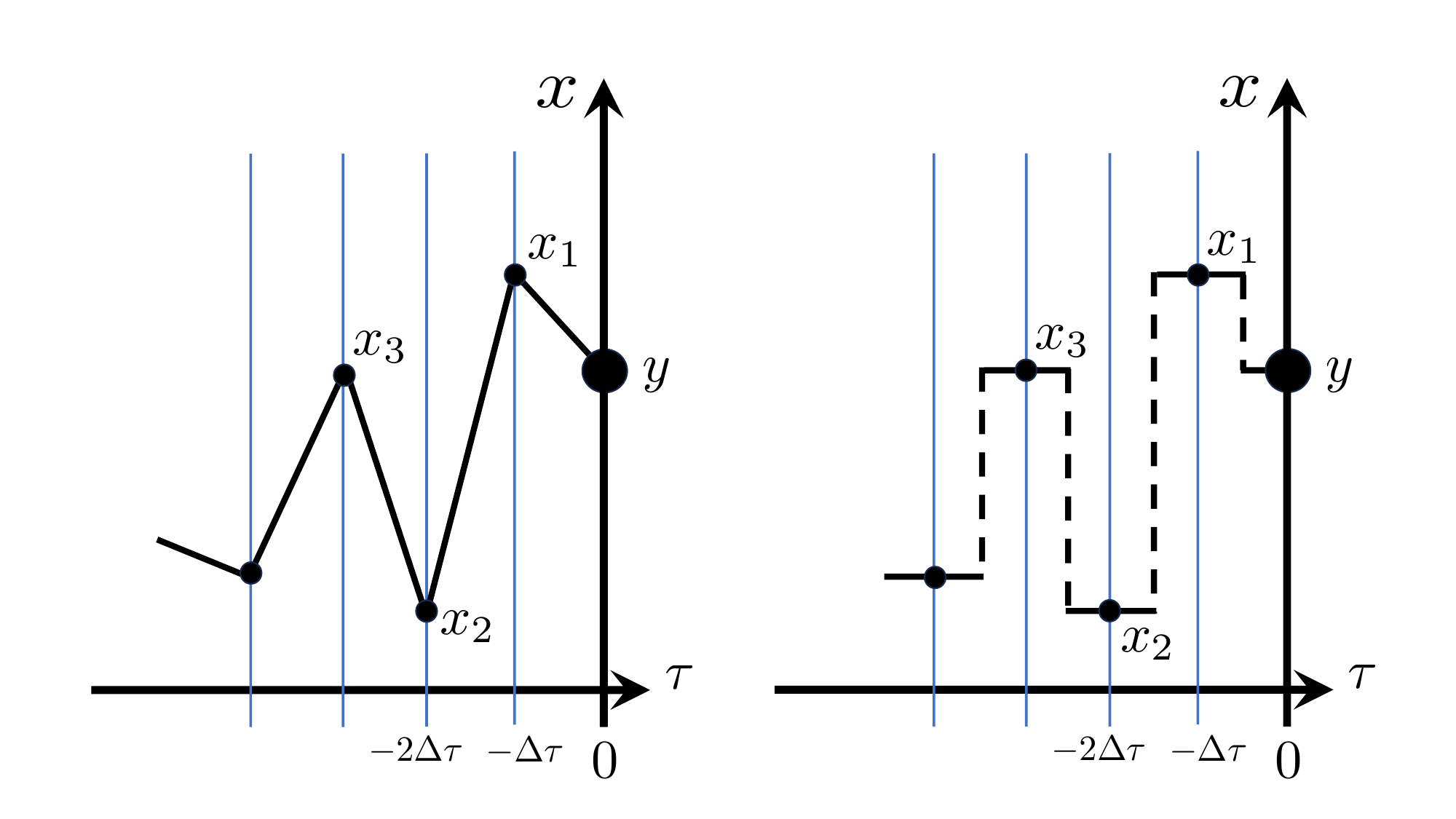}}
\caption{Left: the path-integral in the Euclidean time, denoted by the neural network with ReLU activation, (\ref{xeuc}). Right: the same path-integral with the activation function of the step function, (\ref{xeuc2}).}
\label{fig3}
\end{figure}

Let us start with the popular formula that the path-integral with the Euclidean time can provide the ground state wave function,
\begin{eqnarray}
    \psi(y) \propto
    \int_{x(\tau = -\infty)={\rm any}}^{x(\tau=0)=y} \prod dx_n \exp\left[ -\frac{1}{\hbar}S_{\rm E}\right] \, , 
    \label{psiSE}
\end{eqnarray}
where the Euclidean action is given with the half-line integral over the Euclidean time,
\begin{eqnarray}
    S_{\rm E} = \int_{-\infty}^0 d\tau {\cal L}_{\rm E} \, .
    \label{EucS}
\end{eqnarray}
We parametrize the path-integral configuration $x(\tau)$ with the Euclidean time shown in the left panel of Fig.\ref{fig3}, in the similar manner,
\begin{eqnarray}
    x(\tau) = y + \sum_{n=1}^\infty w_n {\rm ReLU}(-\tau -(n-1)\Delta \tau) \, .
    \label{xeuc}
\end{eqnarray}
Note that we chose the negative sign in the argument of the ReLU 
such that the definition
of the zigzag configuration is backward in the Euclidean time evolution. 
Then the path-integral measure is translated to the language of the neural network parameters through
\begin{eqnarray}
    x_n = y + \Delta \tau\sum_{k=0}^n k w_{n-k+1} \, .
\end{eqnarray}

As an explicit example, using $W_n$ defined earlier in (\ref{Ww}), the Euclidean action for the harmonic oscillator is explicitly evaluated as
\begin{eqnarray}
    S_{\rm E}
    = \sum_{n=1}^\infty
\left[
\frac{m\Delta\tau}{2} (W_n)^2
+ \frac{k \Delta \tau}{2}\left(
y + \Delta \tau \sum_{k=1}^n W_k
\right)^2
\right] \, .
\label{SEHO}
\end{eqnarray}
Since $W_n$ is independent from each other, the measure for $W_n$ is trivial again.
In summary, we find that the ground state wave function has an expression
\begin{eqnarray}
    \psi(y) \propto
    \int \prod_{n=1}^\infty dw_n \exp\left[ -\frac{1}{\hbar}S_{\rm E}(\{w_n\}, y)\right] \, .
\end{eqnarray}
In this way, for any given quantum mechanical Lagrangian, the ground state wave function is written 
as an integral over the network parameters with the probability distribution given by the Euclidean action evaluated with
the neural network output $x(\tau)$.

This Euclidean approach can also 
computes the partition function at finite temperature systems with $\tau^{\rm fin}-\tau^{\rm ini} = \beta \hbar$, with $\beta = 1/T$ being the inverse temperature.


\subsection{Random walk interpretation and stochastic quantization}

The meaning of the neural network parameters can be explicit when we use a specific activation function.
Let us use the step function $\theta(\tau)$ as the activation functions, as before, 
and write 
\begin{eqnarray}
    x(\tau) = y + \Delta \tau \sum_{n=1}^\infty W_n 
    \, \theta\left(-\tau -\left(n-\frac12 \right)\Delta \tau\right) \, ,
    \label{xeuc2}
\end{eqnarray}
instead of (\ref{xeuc}). See the right panel of Fig.~\ref{fig3} for this description. Notice that the bias of the first affine map is shifted by $1/2$ so that the position of the path-integrated coordinate $x_n=x(\tau=-(n-1)\Delta\tau)$ is well-defined,
\begin{eqnarray}
    x_n= y + \Delta \tau \sum_{k=1}^n W_k \,  \,  \, .
    \label{xnyWE}
\end{eqnarray}
Then, the probability distribution for the harmonic oscillator, written in terms of the network parameters $\{ W_n \}$, has the same expression as (\ref{SEHO}).

We like to point out that this form of our neural network with the step function activation allows
an interpretation as a stochastic quantization. 
The stochastic quantization is another method of quantization. 
The method intrinsically uses randomized noise in the Euclidean-time evolution, and thus sounds quite parallel to the statistical average of neural networks. 
Below, we discuss that the use of the step function activation function in the feedforward neural networks leads to the stochastic quantization.

First, let us briefly remind the readers of the stochastic quantization {\it a la} Nelson \cite{Nelson:1966sp} (see App.~\ref{app:B} for a brief review).
The stochastic process is provided by the Langevin equation for the evolution along the Euclidean time $\tau$, which is
\begin{eqnarray}
    \frac{\partial x(\tau)}{\partial \tau}
    = F(x(\tau),\tau) + \sqrt{2D} \, \eta(\tau)
    \label{Lan}
\end{eqnarray}
where $D$ is the diffusion constant and $\eta(\tau)$ is the normalized random Gaussian noise,
\begin{eqnarray}
  \langle \eta(\tau)\rangle =0, \quad
  \langle \eta(\tau_1)\eta(\tau_2)\rangle =
  \delta(\tau_1-\tau_2) \, .
\end{eqnarray}
In the infinite time limit $\tau\to\infty$, because of the repetition of the random kick $\eta$, the resultant $x(\tau=\infty)$ is subject to a certain probability distribution $\rho(x)$. Once the drift term $F(x(\tau))$ is properly chosen, this distribution coincides with the probability distribution of the given quantum system. For example, for the ground state of a harmonic oscillator, the drift term is just the Hooke's law, $F\propto x(\tau)$.

The essential part of the stochastic quantization is that the evolution is given by the random kick $\eta$ which is Markovian, with the drift term. This could be seen alternatively as that the mean of $\eta$ is shifted by the drift $F$ at each fictitious time $\tau$, if one wants to regard the whole right hand side of the Langevin equation (\ref{Lan})
as a random kick. Denoting this mean-shifted $\sqrt{2D}\eta(\tau) + F(x(\tau))$
as $\tilde{\eta}(\tau)$, the discretized version of the Langevin equation (\ref{Lan}) is simply
\begin{eqnarray}
    x_{n+1} = x_{n} + \Delta \tau \, \tilde{\eta}_n
    \label{kick}
\end{eqnarray}
where we have discretized the Euclidean time as $\tau \equiv n \Delta \tau$.

Now, let us turn to our neural network representation (\ref{xnyWE}) with the step function activation. 
It is rewritten as
\begin{eqnarray}
    x_{n+1}= x_{n} + \Delta \tau \, W_{n+1},
    \label{randomw}
\end{eqnarray}
which is basically the same as (\ref{kick}).\footnote{
As we have mentioned in Sec.~\ref{sec:Intro}, 
time evolution of neurons with noise allows a Langevin representation \cite{Somp, Schucker, Helias, Grosvenor:2021eol} which can be formulated with a path integral. However, 
the way the Langevin equation is mapped to the neural networks with noise is different 
from ours. For details, see App.~\ref{app:B}.}
Therefore the backward-time-evolution along $\tau$, discretized, is given by a random kick $\{W_n\}$. This is a random walk, and the stochastic quantization. Therefore we reach the picture that the parameters of the neural network with the step function activation amounts to the random noise used in the stochastic quantization.

The precise relation between the force field $F$ in the Langevin equation 
and the corresponding quantum mechanical action in the stochastic quantization is rather technically involved, and we provide it in App.~\ref{app:B}.

\begin{figure}[t]
\centerline{\includegraphics[width=0.8\columnwidth]{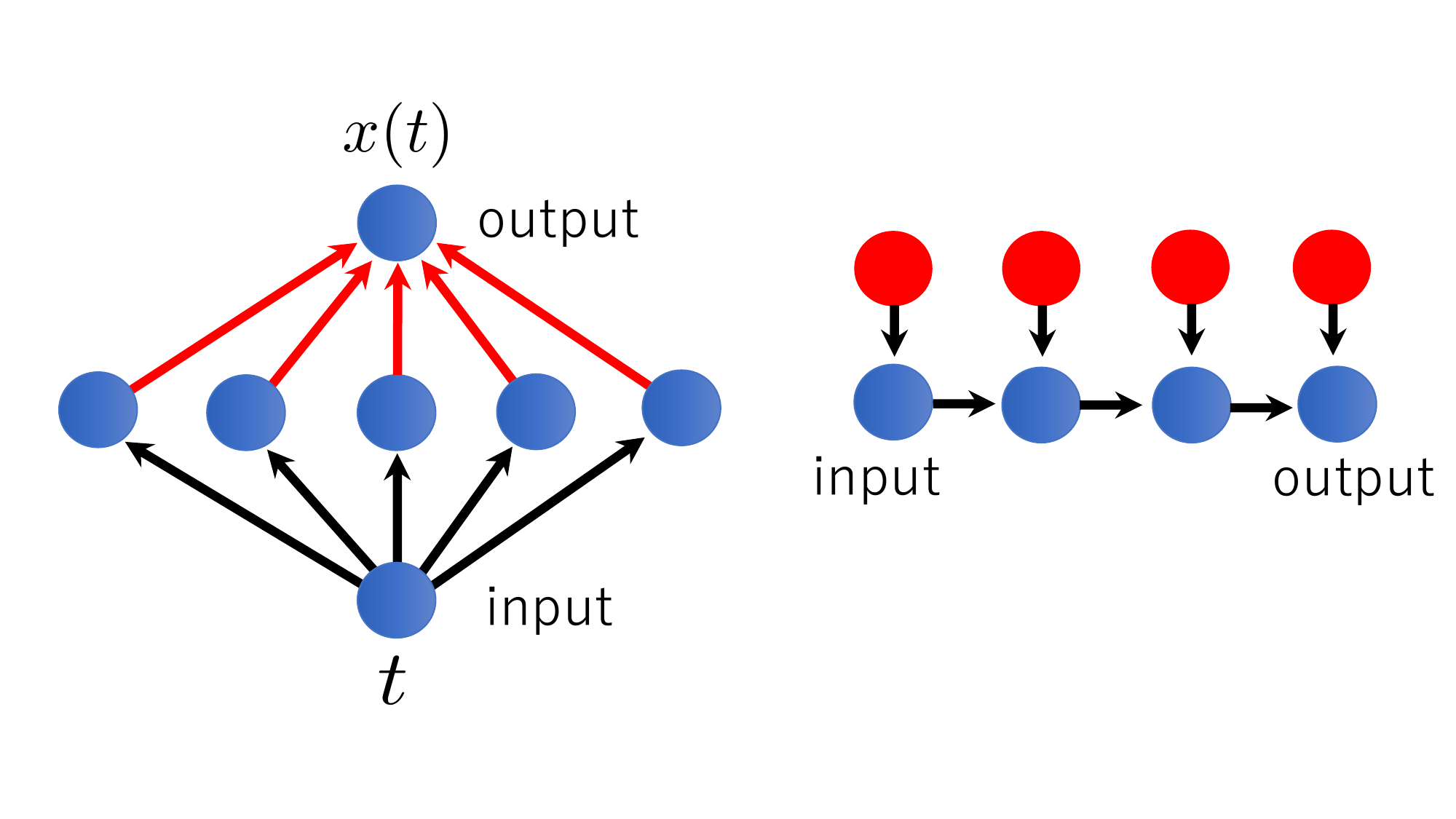}}
\vspace{-10mm}
\caption{Left: our neural network architecture (the network parameters in the red arrows are statistically distributed). Right: the discretized Langevin equation (the red circles represents random units).}
\label{fig5}
\end{figure}

In the right panel of Fig.~\ref{fig5}, we draw a schematic picture for the neural network for the 
stochastic quantization. The time evolution by the Langevin equation is discretized, with the random kink at each time. In the figure, the input is the initial $x(\tau_0)$ and the output is the final $x(\tau_N)$. The time evolution is from left to right, and the red disks depict the random units. This neural network architecture is shared with popular recurrent neural networks and diffusion models. The architecture is compared with our original neural network architecture which is shown in the left panel of Fig.~\ref{fig5}, in which the red arrows depict the random parameters. As can be seen, the orientation of the flows of the information are different, but the orientation of the time evolution is shared. In the sense that these two architectures result in representing the same phenomena, this is a duality of the neural networks.

\section{Summary and discussions}

In this paper, we have shown an equivalence between a generic quantum mechanical system and a statistical neural network, by providing a map between a quantum path-integral
of a single particle and a single-hidden-layer perceptron with infinite width in which the network parameters follows a specific distribution determined by the quantum mechanical action.
This map provides a new look at the relation between quantum physics and machine learning --- the way the universe works and the way the universe is learnt.

The essence of the map is that it identifies the arbitrary zigzag configuration of $x(t)$ in the path-integral definition of quantum mechanics as a neural network output with the ReLU activation function.
This is thought of as a part of the universal approximation theorem for the neural network (note that the map is an exact representation, not an approximation). In the course of the identification, the physical meaning of the neural network parameters becomes evident: the path-integration not by the original $x(t)$ but by the velocity $\dot{x}(t)$
or by the acceleration $\ddot{x}(t)$ (see Table \ref{tab1}).

As we have mentioned in Sec.~\ref{sec:Intro}, the neural network field theory (NNFT) \cite{Halverson:2020trp,Halverson:2021aot,Demirtas:2023fir} provides an intriguing new representation of QFTs and has motivated us to perform the present study.
As described in Sec.~\ref{sec:ACT}, we may understand roughly the NNFT and our map differ in the choice of the activation function. However, looking at the details, we may find more differences, 
in the treatment of time direction and in the interpretation of the statistical neural networks. 
The NNFT does not fully use the network biases while in our case 
the network biases are crucial to get the path integral expression with discretized time direction.
In addition, our path-integral interpretation works only at the strict 
large $N$ limit, while NNFT admits an interpretation of the $1/N$ corrections.
In our case, a finite $N$ would give rise to
discretization errors due to the finite temporal lattice unit.
Another difference is that for the representation of QFTs, the neural network input in our case is the value of time $t$, while in the NNFT it is the full spacetime coordinates.
To find out how general the similarity between neural networks and quantum mechanics is, it would be beneficial to explore deeper connections between these two approaches.

We use the path integral to represent quantum systems, which is a distinction from other approaches. 
Here we list some merits of our path-integral formulation. 
First, the path-integral quantization does not need the Gaussian limit. In fact, 
intrinsically interacting system such as sigma models and spin systems\footnote{For spin systems one can use the coherent state path-integral formulation (note a careful treatment in the discretization, \cite{Shibata:1998ht}).} can be formulated as a neural network.
Second, path-integral quantization can be applied also to some non-local theory (see the book by Feynman and Hibbs \cite{FeynmanHibbs} for an example).\footnote{In {\it conclusions} of the book by Feynman and Hibbs \cite{FeynmanHibbs}, it says: {\it ``We can only say that if the problem is not Gaussian, it can at least be formulated and studied by using path integrals. ... The only example of a result obtained with path integrals which cannot be obtained in simple manner by more conventional method is the variational principle.''} The variational principle here refers to the method to approximate ground state energy by using an effective action whose coefficient of each term can be variationally tuned.} 
Third, the change of dynamical variables is understood intuitively in the path-integral quantization. The dualities in field theories and holographic principles are formulated well upon the path-integral formulation.
These would be just some examples of the merits of the path-integral method, and we expect that our map would gain the benefit. Note that, although our formulation is quite generic, it is still limited to bosonic scalar field theories. The treatment of fermionic fields and gauge fields remains a challenge, in view of incorporation of Grassmann variables and gauge volumes in our neural network formulation.

As we have described at the end of Sec.~\ref{sec:Euclid}, the stochastic quantization allows a Langevin-like equation whose time-discretization gives another neural network architecture for time evolution which
has been studied earlier in \cite{Somp,Schucker,Helias,Grosvenor:2021eol}. Since it results in the same quantum mechanics as ours, we find a duality.
In general, it is known that the use of the extra dimension (in the present case, the Euclidean time direction to produce ground state of quantum systems) can allow several different but equivalent formulations. As an example in the context of neural networks, the deep Boltzmann 
machines used for quantum mechanical ground state \cite{Carleo:2017nvk,Imada} may allow physical interpretation 
of Euclidean time evolution. The deep Boltzmann machines have statistical average over hidden unit values, while that can be traded with a statistical average over network parameters equivalently (see \cite{Aarts:2023uwt} for a decomposition of network weight matrices in a QFT-like formulation of the Boltzmann machine). The duality would provide a further connection to quantum gravity \cite{Hashimoto:2019bih,You:2017guh,Hashimoto:2018ftp,Hashimoto:2018bnb} where gravity and matter are represented by network parameters and network units respectively, while both need to be summed as a path integral.

We have discussed 
the natural meaning of the network parameters in Sec.~\ref{sec:ACT}, and through the argument given in App.~\ref{app:A} for the generality of the neural network we use,
we find a possibility that generic wide 3-layer perceptrons would allow a quantum mechanical interpretation. Revealing any significance of this statement that connects generic neural networks and generic quantum mechanics would deserve further research, for a hidden mechanism of how we perceive the world.


\section*{Acknowledgments}
The authors would like to thank Gert Aarts, David Berman, James Halverson, Anindita Maiti, Akiyoshi Sannai and Daichi Takeda for valuable discussions. The authors appreciate String Data 2023 conference
and Machine Learning Physics 2023 Kyoto conference for their hospitality, 
during which a part of this paper was discussed.
The work of K.~H.~is supported in part by 
JSPS KAKENHI Grant Nos.~JP22H01217, JP22H05111, and JP22H05115.
The work of Y. H. was supported in part by JSPS KAKENHI Grant No. JP22H05111.
The work of J.~T.~Y.~is supported in part by JSPS KAKENHI Grant Nos.~JP23KJ1311.


\appendix


\section{Generality of our network via gauge symmetry}
\label{app:A}

In this appendix, we show that our assumptions on the network parameters can be relaxed
so that the neural networks are generic, for the correspondence with quantum mechanical systems.
Remember that 
the neural network architecture shown in Fig.~\ref{fig1} in Sec.~\ref{sec:QMRNN}
admits specific values for the network parameters $\tilde{w}_n$ 
and $\tilde{b}_n$, which are given in (\ref{tilfix}). 
To relax the condition (\ref{tilfix}), we use a gauge symmetry of the neural network
and the arbitrariness of the time division of the path integral.

First, notice that (\ref{xtrelu}) can be written, for positive $\tilde{w}_n$, as follows: 
\begin{eqnarray}
x(t) = \sum_{n=0}^{N-1} w_n \tilde{w}_n {\rm ReLU}(t + \tilde{b}_n/\tilde{w}_n) + b \, .
\label{xtrelu3}
\end{eqnarray}
This is because the activation function ReLU$(x)$ enjoys the scaling relation
\begin{eqnarray}
{\rm ReLU}(\alpha x) = \alpha {\rm ReLU}(x)
\end{eqnarray}
for $\alpha\geq 0$. If we regard the combination $w_n \tilde{w}_n$ as a new parameter, it amounts to  
putting $\tilde{w}_n=1$.
Another way to look at this is a gauge fixing\footnote{See \cite{Hashimoto:2024rms} for the interpretation of the gauge symmetry as a spacetime diffeomorphism.}; 
there exists a gauge symmetry of a neural network that
the input-output relation is invariant under the transformation 
\begin{eqnarray}
w_n \to \alpha^{-1} w_n \, , \quad 
\tilde{w}_n \to \alpha \tilde{w}_n \,  \, , \quad 
\tilde{b}_n \to \alpha \tilde{b}_n \, ,
\end{eqnarray}
for any fixed $n$. Using this gauge degrees of freedom, we can fix the gauge,
with the gauge condition $\tilde{w}_n=1$.

Then the remaining degrees of freedom is $\tilde{b}_n$ in the expression
\begin{eqnarray}
x(t) = \sum_{n=0}^{N-1} w_n  {\rm ReLU}(t + \tilde{b}_n) + b \, .
\label{relu3}
\end{eqnarray}
If we make an integration over this $\{\tilde{b}_n\}$ with a trivial measure,
then a given configuration $\{\tilde{b}_n\}$ in the integrand corresponds to
having an arbitrary discretization of the $t$ axis.
(The order of $\{\tilde{b}_n\}$ does not cause a problem; we can rearrange
the index $n$ such that $\{\tilde{b}_n\}$ aligns in a decreasing fashion.
The integration measure $\prod d\tilde{b}_n$ is divided by the volume of the 
permutation group, $N!$.)
There could be two ways to gauge-fix $\tilde{b}_n \to -t_{n-1}$.
First is to resort to the expectation that the path-integral itself may be free of
the discretization manner, once properly defined. Lattice artifact should go away in
the continuum limit $N\to\infty$.
The other way is to bring this arbitrary discretization to the standard form (\ref{xn})
by assuming a time-reparametrization invariance for the particle action $S$.
For example, a relativistic action
\begin{eqnarray}
S = \int dt \, \sqrt{(\dot{X}^0)^2 - (\dot{X}^1)^2}
\end{eqnarray}
admits the time reparametrization invariance, thus any choice of $\{\tilde{b}_n\}$
may be rearranged, and we may gauge-fix $\tilde{b}_n \to -t_n$. In any case, we expect 
that the integration over possible $\{\tilde{b}_n\}$ is a redundancy.

To summarize, the fixing condition for the network parameters $\tilde{w}$ and $\tilde{b}$ 
given in (\ref{tilfix}) can be relaxed to have the general expression (\ref{xtrelu})
while the equivalence to the quantum mechanics is maintained.
One can even write the probability distribution for the network parameters with using
the general (\ref{xtrelu}), although it looks more complicated than the one 
for the gauge-fixed version which we saw in this paper.


\section{A brief review of stochastic quantization}
\label{app:B}

In this appendix we provide a brief review of the stochastic quantization introduced by Nelson
\cite{Nelson:1966sp}.
There are two primary approaches in stochastic quantization.
One is by Nelson \cite{Nelson:1966sp}, in which the wave function $\psi(x)$ for ground states and excited states is obtained through a stochastic motion of a particle $x(\tau)$,
where $\tau$ is a stochastic time. 
The other approach is by Parisi and Wu \cite{Parisi:1980ys}, in which configurations $x(t)$ following the given probability density are produced by a stochastic ``motion'' $x(t,\tau)$. 
While the latter method is widely used especially in lattice quantum field theories, we here focus on the former.
This is because the latter approach introduces 
both a real time $t$ and a stochastic time $\tau$, which 
does not suits our neural network architecture straightforwardly.
For a comprehensive review of stochastic quantization, 
see \cite{Damgaard:1987rr}.

We start with the following Langevin equation
\begin{eqnarray}
    \frac{dx(\tau)}{d\tau}= F(x(\tau)) + \sqrt{2D} \,  \eta(\tau)
    \label{Lang}
\end{eqnarray}
where $\tau$ is the stochastic time, 
$F(x(\tau))$ represents a deterministic force, 
and stochasticity is given by a Gaussian white noise represented by $\eta(\tau)$
with strength $D$.
The noise field is normalized so that
\begin{eqnarray}
    \langle \eta(\tau) \eta(\tau')\rangle = \delta(\tau-\tau') \, ,
\end{eqnarray}
where $\langle \cdot \rangle$ denotes the average over noise realizations. 
Suppose one starts with $\tau=-\infty$ at some value of $x(-\infty)$ and let it develop according to the Langevin equation (\ref{Lang}) till time $\tau=0$. 
When the system has a stable stationary distribution, 
the probability distribution at $x(\tau=0)$ is 
independent of the initial condition $x(-\infty)$ and
is given by the following path-integral expression,
\begin{eqnarray}
    P(x) = {\cal N} \int {\cal D}x \, \exp \left[
    -\int_{-\infty}^0 d\tau \, L^{(\rm s)}
    \right]
    \label{probP}
\end{eqnarray}
where ${\cal N}$ is a normalization constant,
and $L^{(\rm s)}$ is a ``stochastic Lagrangian'' given by
\begin{eqnarray}
    L^{(\rm s)} \equiv 
    \frac{1}{4D}
    \left(
     \frac{dx}{d\tau} 
     - F(x)
    \right)^2
    + \frac12 \frac{\partial F(x)}{\partial x} \, ,
    \label{eq:s-lag-1}
\end{eqnarray}
where we have used the Stratonovich product 
for the cross term from the first term of Eq.~\eqref{eq:s-lag-1}, meaning that ``$\int F(x) dx$'' should be understood as
\begin{equation}
- \frac{1}{2D} 
\int F(x) d x 
\equiv 
- \frac{1}{2D} 
\int 
\frac{1}2 
\left(
F(x(\tau)) + 
F(x(\tau + d\tau ))
\right)
(x(\tau + d \tau) - x(\tau)) \, . 
\label{eq:cross-term}
\end{equation}
In other words, we employ the midpoint prescription.
This choice is made for later convenience and 
the discussion below does not depend on this choice.
Note that, while the individual terms of Eq.~\eqref{eq:s-lag-1} depend on the choice of prescriptions, the whole Lagrangian is independent of it. 
Note also that this $L^{(\rm s)}$ is {\it not} the Lagrangian of the quantum mechanics which gives the probability distribution $P(x)$ for the ground state, as we will see below: it differs by the term \eqref{eq:cross-term}.

Let us here consider the case where $F(x)$ is a potential force, {\it i.e.} there exists a function $G(x(\tau))$ such that 
\begin{eqnarray}
    F(x) = -\frac{dG(x)}{dx} \, .
    \label{FG}
\end{eqnarray}
Then we find
\begin{equation}
- \frac{1}{2D} 
\int_{- \infty}^0 
F(x) d x 
= 
\frac{1}{2D} 
\int_{- \infty}^0 
\frac{d G (x)}{d x} 
 d x 
 = 
 \frac{1}{2D} G(x(0))  + {\rm const.} ,
\end{equation}
where we used the fact that 
in the case of Stratonovich product
the usual chain rule holds, 
$dG =  (dG/dx) dx$.
Substituting this to (\ref{probP}), we find
\begin{eqnarray}
    P(x) = &{\cal N}' 
    \exp\left[-\frac{1}{2D}G(x(0))\right]
    \nonumber \\
    &
    \times \int {\cal D}x \, \exp \left[
    \frac{-1}{4D}\int_{-\infty}^0 d\tau \, \left(
     \left(\frac{dx}{d\tau}\right)^2
    + \left(\frac{\partial G}{\partial x}\right)^2
    -2D \frac{\partial^2 G}{\partial x^2}
    \right)
    \right] \, .
    \label{probP2}
\end{eqnarray}
Identifying $2D = \hbar/m$
and defining
\begin{eqnarray}
    {\cal L}_{\rm E} \equiv 
    \frac{m}{2}\left(\frac{dx}{d\tau}\right)^2 +\frac{m}{2} 
    \left(\frac{\partial G}{\partial x}\right)^2
    -\frac{\hbar}{2}\frac{\partial^2 G}{\partial x^2} \, , 
    \label{LEQM}
\end{eqnarray}
we find the expression
\begin{eqnarray}
    P(x) = {\cal N}' 
    \exp\left[-\frac{m}{\hbar}G(x(0))\right]
    \int {\cal D}x \, 
    \exp \left[
    \frac{-1}{\hbar}\int_{-\infty}^0 d\tau \, {\cal L}_{\rm E} 
    \right] \, .
    \label{probP3}
\end{eqnarray}
As we will see, ${\cal L}_{\rm E}$ is the Euclidean Lagrangian of the quantum mechanics whose ground state probability distribution is given by $P(x)$.

Note that the second factor of the RHS of \eqref{probP3} 
is nothing but the ground state wave function $\psi(x)$. 
In order for $P(x)$ on the LHS of (\ref{probP3}) 
to satisfy the following relation
\begin{eqnarray}
    P(x)=|\psi(x)|^2 \,, 
\end{eqnarray}
we need to have
\begin{eqnarray}
    \exp\left[-\frac{m}{\hbar}G(x(0))\right]
    = {\cal N}''
    \int {\cal D}x \, 
    \exp \left[
    \frac{-1}{\hbar}\int_{-\infty}^0 d\tau \, {\cal L}_{\rm E} 
    \right] \, .
    \label{probP4}
\end{eqnarray}
This is indeed true, as we see below.
We look at the Euclidean Lagrangian (\ref{LEQM}) and bring it to the Lorenzian Lagrangian to find the Lorenzian Hamiltonian
\begin{eqnarray}
    {\cal H} \equiv 
    \frac{1}{2m}p^2+\frac{m}{2} 
    \left(\frac{\partial G(x)}{\partial x}\right)^2
    -\frac{\hbar}{2}\frac{\partial^2 G(x)}{\partial x^2}\, .
    \label{LEQMH}
\end{eqnarray}
Then, one can explicitly check that the LHS of \eqref{probP4} is the zero-energy solution of the Schr\"odinger equation,
\begin{eqnarray}
    {\cal H}\exp\left[-\frac{m}{\hbar}G(x)\right] =0\, .
\end{eqnarray}
Thus, we have shown that the stationary distribution 
of the Langevin dynamics (\ref{Lang}) is related to the ground state of the quantum mechanical system with the Euclidean Lagrangian (\ref{LEQM}) through
the identification $2D = \hbar/m$.


Concerning our intention to relate quantum mechanics and neural networks, we have two important comments. 
\begin{itemize}
    \item 
First, as we have mentioned, the path-integral expression 
(\ref{probP}) of the Langevin dynamics is not equivalent to the path-integral expression (\ref{psiSE}) for the ground state. There is an important prefactor (\ref{probP4}) to make the wave function squared. So, the meaning of the ``path" in (\ref{probP}) is different from
what we study in this paper.
    \item 
Second, the obtained quantum mechanical Hamiltonian includes $\hbar$, as seen in the last term of (\ref{LEQMH}). 
This can be thought of as an effective Hamiltonian 
that includes quantum corrections.
Alternatively, this $\hbar$ term is thought of as an ambiguity of the ordering of operators; in fact, in terms of operators, the Hamiltonian is written as
\begin{eqnarray}
    {\cal H} = 
    \frac{1}{2m}\hat{p}^2+V(\hat{x}) - \frac{i}{\sqrt{2m}}\left[\hat{p},\sqrt{V(\hat{x})}\right]\,  \, ,
    \label{LEQMH2}
\end{eqnarray}
with $V(x) \equiv (m/2)(dG/dx)^2$.
So, if one wants to start with a classical Hamiltonian and quantize it, one needs to be careful in this $\hbar$-dependent term.
\end{itemize}
These subtle complications in fact prevent us from relating the neural network representation of the Langevin dynamics to quantum mechanics directly.


\end{document}